\documentclass[usenatbib]{mn2e}
\usepackage{epsf}

\title{Dark matter halo response to the disk growth}

\begin{document}
\author[J.-H. Choi et al.]{Jun-Hwan Choi, Yu Lu, H. J. Mo \& Martin D. Weinberg \\
Department of Astronomy, University of Massachusetts, Amherst, MA 01003 \\  
jhchoi@nova.astro.umass.edu, luyu@astro.umass.edu, 
hjmo@nova.astro.umass.edu, weinberg@astro.umass.edu}
\maketitle

\begin{abstract}
  We consider the sensitivity of the circular-orbit adiabatic
  contraction approximation to the
  baryon condensation rate and the orbital structure of dark matter
  halos in the $\Lambda$CDM paradigm.  Using one-dimensional
  hydrodynamic simulations including the dark matter halo mass
  accretion history and gas cooling, we demonstrate that the adiabatic
  approximation is approximately valid even though halos and disks may
  assemble simultaneously.  We further demonstrate the validity of the
  simple approximation for $\Lambda$CDM halos
  with isotropic velocity distributions using
  three-dimensional N-body simulations.  This result is easily
  understood: an isotropic velocity distribution in a cuspy halo
  requires more circular orbits than radial orbits.  Conversely, the
  approximation is poor in the extreme case of a radial orbit halo. 
  It overestimates the response a core dark matter halo, where radial 
  orbit fraction is larger.  
  Because no astronomically relevant models are dominated by
  low-angular momentum orbits in the vicinity of the disk and the
  growth time scale is never shorter than a dynamical time, we
  conclude that the adiabatic contraction approximation is
  useful in modeling the response of dark matter halos to the 
  growth of a disk.   
\end{abstract}


\section{Introduction}
\label{sec:intro}

In hierarchical structure formation, galaxies form in
gravitationally collapsing dark matter halos.  The 
dissipative baryonic matter cools and condenses in dark matter halo
\citep{wr78}.  \citet{blumenthal86} described the halo response to
this condensation assuming a spherical profile with circular orbits
and adiabatic disk growth \citep{bw84}.  For an adiabatic change, the angular momentum is
invariant: $J^2\propto rM(r)=\hbox{constant}$. Then, given the
distribution of the baryonic disk $M_d(r)$ and the initial dark matter
distribution $M_i(r)$, the final distribution of dark matter $M_f(r)$
must satisfy
\begin{eqnarray}
  &M_f(r_f)r_f =  M_i(r_i)r_i&  \nonumber \\
  &M_f(r_f) =  M_d(r_f)+M_i(r_i)(1-m_d)&
\label{eq_ad}
\end{eqnarray}
where $m_d$ is mass fraction of the disk.  Further studies of this
model include \citet{ryden88, ryden91}, and \citet{flores93}.  Recently
\citet{jnb02} find that the adiabatic contraction approximation is in
agreement with their simulations.

\begin{figure*}
\setlength{\epsfxsize}{0.75\textwidth}
\centerline{\epsfbox{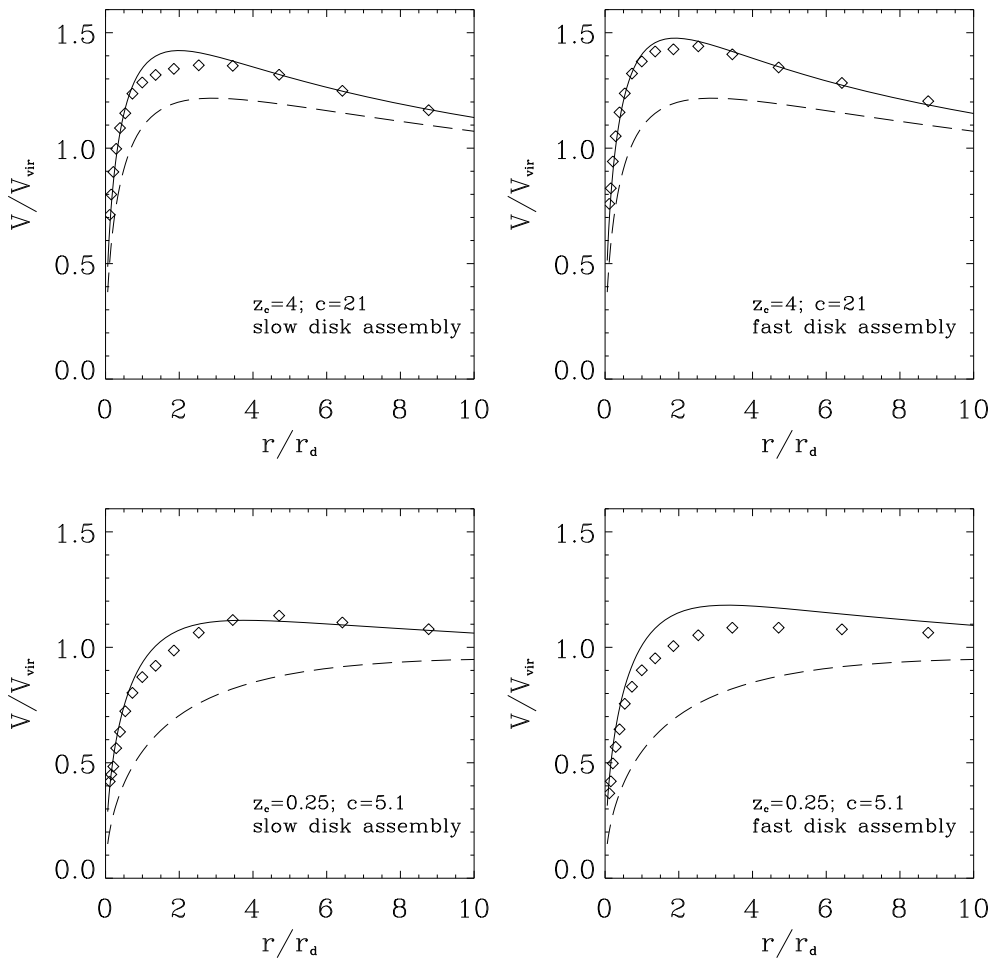}}
\caption { The rotation curves of simulated halos with
different formation time and gas cooling phases compared with the
rotation curves with the predictions of adiabatic contraction
approximation.  The upper-left panel shows the rotation curves for
early-formed halo ($z_c=4$) with normal cooling (hot mode).  The
upper-right panel is for early formed halo ($z_c=4$) with artificially
high cooling (cold mode).  The bottom-left panel is for late formed
halo ($z_c=0.25$) with normal cooling, and the bottom-right panel is
for late formed halo ($z_c=0.25$) with high cooling.  In each panel,
the diamonds denotes simulation results, the solid lines show the
predictions of adiabatic contraction approximation, and the dashed
lines show the rotation curves of the control model, in which cooling
and baryon condensation are turned off.  }
\label{fig:MAH}
\end{figure*}

Despite the wide usage of the adiabatic contraction approximation,
discrepancies between observations and theoretical predictions
motivate a detailed check of its validity.  First,
\citet{blumenthal86} simply assumes that the disk growth time is much
longer than the dark matter halo dynamical time.  However, individual
halos may grow simultaneously with their disks and have different assembly
histories \citep{We02, zh03, li05}.  Cosmological simulations of the
galaxy formation have shown that gas accretion in CDM halos proceeds in
two distinct modes depending on the mass of the halo \citep{keres05,
bd03}.  In massive halos, gas accretion is dominated by ``hot mode''
in which gas is first heated up to the virial temperature of the halo
and then cools to settle gradually into the halo centre.  By contrast,
gas accretion into small halos is dominated by ``cold mode'' in which
cold gas sinks in a dynamical time without being shock heated.  These
scenarios may affect the validity of the adiabatic contraction
formula.  Secondly, dark matter is not arranged on circular orbits and
therefore $rM(r)$ is not strictly conserved.  \citet{barnes87}, \citet{sell99}, 
and \citet{SM05} report that the approximation overestimates the contraction
measured in simulations.
Moreover, recently \citet{Gnedin04} claimed similar findings in a
cosmological simulation.  These authors suggest that the discrepancy
is due to the circular orbit assumption.  

In this paper, we use idealised numerical
experiments to investigate the effect of the two assumptions in the
adiabatic contraction approximation (eq. \ref{eq_ad}) and provide
physical intuition for the numerical trends.  In \S\ref{sec:timing},
we use one-dimensional simulations which incorporate dark matter halo 
mass accretion history as well as gas cooling to test the adiabatic 
disk growth assumption in realistic forming halo. We find that disk 
growth time scale is always longer than dynamical time of the halo 
in many cases. We also show that the
continued dark-matter mass accretion has little affect in the inner halo.  
In \S\ref{sec:halo}, we test 
the circular orbit assumption using high resolution N-body simulations
with cosmological dark matter halo initial conditions.  As expected,
we find that radial orbits reduce the dark matter halo response to
disk growth predicted by the simple circular-orbit approach. However, 
in order to maintain an
isotropic velocity distribution in cuspy halos, a circular population
is much larger than radial orbit population.
This explains the often-observed consistency between
simulations of a dark-matter cuspy halo to disk growth.  
We study typical CDM halos in \S\ref{sec:typical} and summarise in
\S\ref{sec:summary}.

\section{Adiabatic contraction during CDM halo formation}
\label{sec:timing}

We investigate a dark matter halo response to disk growth for several
different halo and disk growth time scales using one-dimensional
hydrodynamic simulations \citep[see Appendix A and ][]{lu06}.
We use $5\times10^4$ equal-mass shells to represent the 
dark matter distribution and 500 equal-mass shells to represent 
the gas distribution. The evolution of every gas shell is followed 
until it cools below $10^4$K; thereafter its mass is assigned to 
an exponential disk. The scale-length of the 
disk, $r_d$, is fixed to ${0.05 \over \sqrt{2}}r_{\rm vir}$, where 
$r_{\rm vir}$ is the virial radius of the halo at $z=0$ \citep{MMW98}.  
The disk is assumed to be a rigid exponential disk.
  
We examine both an early- ($z=4$) and late- ($z=0.25$) time
halo-formation scenarios with $M=10^{12}\>{\rm M_{\odot}}$.  
Because the virial mass of the halo is
fixed at $10^{12}\>{\rm M_{\odot}}$, gas accretion is primarily in
the ``hot mode'' phase for this halo mass \citep{keres05}.  The gas is heated to the
virial temperature and then slowly cools to form a disk.  In order
to examine the adiabatic contraction for the ``cold mode'' gas
accretion, we artificially increase cooling rate by a factor of 100.
Then, the accreted gas cools rapidly and joins the disk without shock
heating.  In this case, the disk and its host halo have similar growth
times. 

For the four simulations, two redshifts and two accretion models, 
we measure the rotation curves at the 
present time, which are shown in Figure \ref{fig:MAH}.  
Since disk growth time is longer than dynamical
time of the dark matter halo in all four, the adiabatic contraction approximation
adequately predict the response of dark matter halos.  Although the
host halo continues to accrete dark matter, this accretion primarily
affects the outer halo. In summary, the adiabatic contraction formula
gives an acceptable approximation for these standard scenarios.

\section{Adiabatic contraction for a non-circular orbit distribution}
\label {sec:halo}

\begin{figure}
\centerline{
\epsfxsize=1.0\columnwidth
\epsfbox{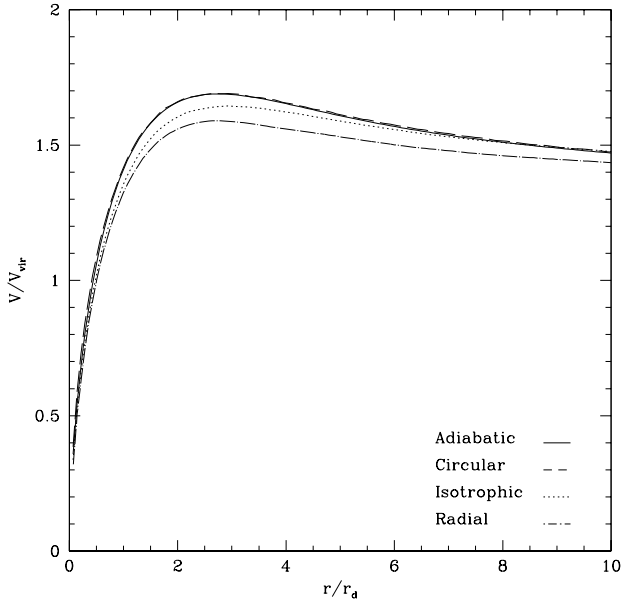}
}
\caption {The final rotation curve for a circular orbit (long-dashed), an
isotropic (dotted), and a radial-orbit dominant (dot-dashed) distribution.  
The dark matter halo is c=12 NFW halo and disk is fiducial disk.  The solid line
is for the adiabatic contraction approximation prediction. 
The rotation curve for the adiabatic contraction
approximation prediction and the circular orbit halo case are almost
identical.  }
\label{fig:Orbit}
\end{figure}

\begin{figure}
\centerline{
\epsfxsize=1.0\columnwidth
\epsfbox{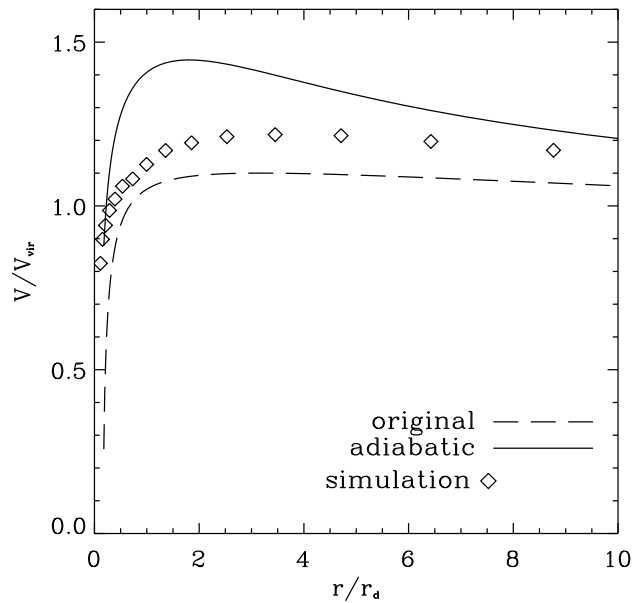}
}
\caption { 
The rotation curve of the pure radial orbit halo. The diamonds 
denote the simulation result, the solid line denotes the prediction 
of adiabatic contraction, and the long dashed line denotes the 
rotation curve of the reference halo. 
}
\label{fig:radial}
\end{figure}

We explore disk growth in a dark matter halo represented by a \citet[hereafter NFW]
{NFW97} profile, $\rho \propto {1}/{[r(r+r_{s})^{2}]}$ with virial radius
$r_{vir}$.  We assume a concentration $c=r_{vir}/r_{s}=12$ consistent
with the rotation curve for a large spiral galaxy and examine three
cases of different anisotropy to study the effects of radial orbits on
the circular-orbit adiabatic prediction: a circular-orbit halo, an
isotropic halo, and a radial orbit biased halo.  
The NFW halo with pure circular orbits is constructed
by assigning each of the particles at radius $r$ a tangential
velocity, $v_{c} = \sqrt{GM(< r)/r}$ in a random tangential direction.
The isotropic and radially biased distribution functions are computed
using the Osipkov-Merritt model \citep{BT87,Osipkov79,Merritt85}
which controls the orbital structure by an anisotropy radius
$r_{a}$.  For an isotropic halo, $r_a=\infty$.  For the
radially biased case, we choose the minimum value for $r_a$ that
results in positive density.  The anisotropy profile in this model
closely corresponds to the profile from a virialized collapse \citep{ENF98, CKK00}.  
The $N=10^6$-particle phase spaces are realized by
a Monte Carlo procedure.  Our rigid exponential disk has mass
$m_{d}=0.04 M_{vir}$ and the disk scale length is $r_{d}=0.014
r_{vir}$ motivated by galaxy formation in $\Lambda$CDM cosmogony
\citep{MMW98,KZS02}.  To mimic disk growth, we increase the disk mass
from zero, keeping the disk scale length unchanged.  To ensure the
validity of the adiabatic approximation, the time scale of disk mass
growth is 10 times longer than the dynamical time of the dark matter
halo at the disk scale length.  The gravitational force on each dark
matter particle is calculated using the self-consistent field code 
(SCF) \citep{cbrock72,cbrock73,ho92,weinberg99}, which
solves Poisson's equation using a set of density-potential
bi-orthogonal function expansions.

\begin{figure*}
\setlength{\epsfxsize}{0.75\textwidth}
\centerline{\epsfbox{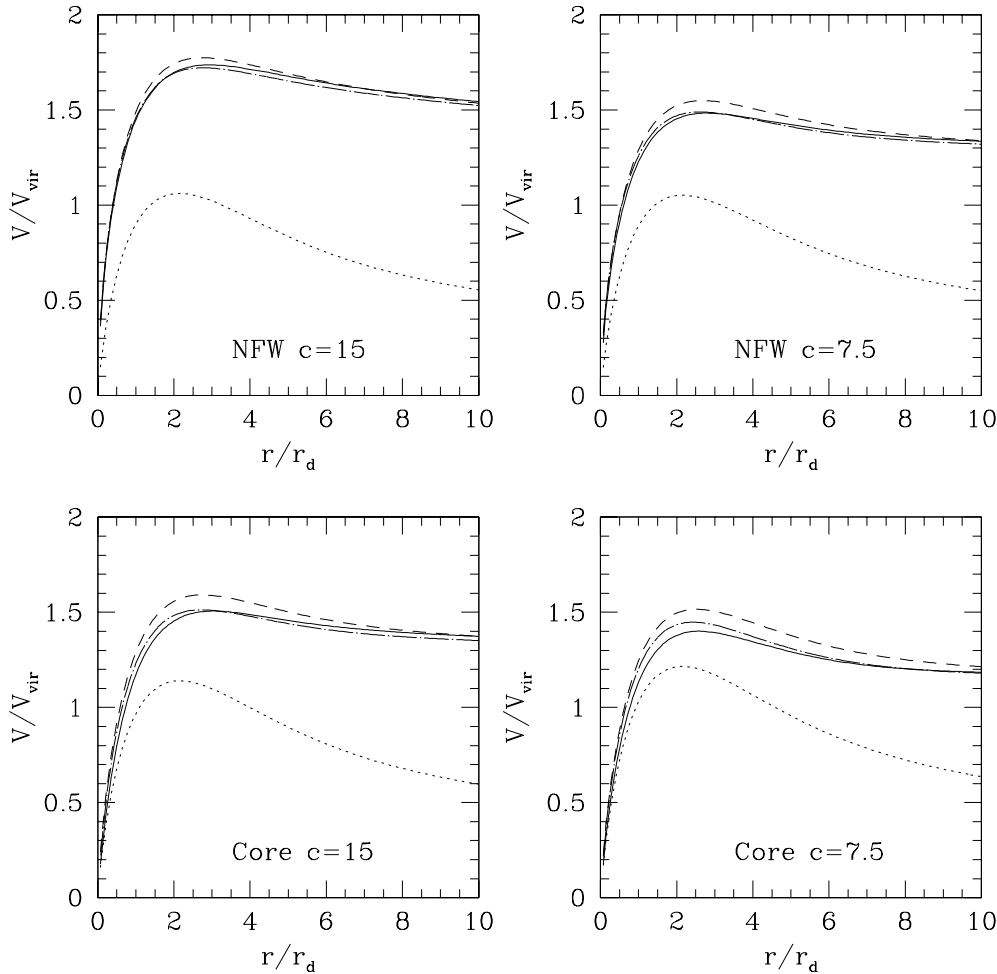}}
\caption {Rotation curves from simulations cooperated with the
adiabatic contraction approximation for isotropic NFW and core halos. 
The dotted line is the rotation  curve from disk only, the solid 
line is the total (disk + dark matter halo) rotation curve from 
the simulation results, and the long dashed line is the total rotation 
curve from the adiabatic contraction approximation.
We also plot the total rotation curve predicted 
by the modified adiabatic contraction approximation 
of \citet{Gnedin04} with $A=0.85$ and $w=0.8$ (dot-dashed line).   
The disk has the fiducial parameters: 
$r_{d}=0.014r_{vir}$ and $m_{d}=0.04m_{vir}$. The overestimation by 
the adiabatic contraction approximation increases as halo concentration 
decreases. The modified adiabatic contraction approximation provides 
a much better fit to the simulation results. 
}
\label{fig:Vhalo}
\end{figure*}

Figure \ref{fig:Orbit} shows the total rotation curves after disk
contraction for the circular orbit case (long dashed curve), the
isotropic case (dotted curve) and the radial orbit biased case
(dot-dashed curve).  For comparison, we also show the prediction of
the adiabatic contraction approximation (solid curve).  The adiabatic
contraction approximation is nearly exact for
circular orbits, but overestimates the contraction when the eccentric
orbit contribution increases. The radial orbit biased halo model is not 
significantly biased toward radial orbit around disk ($r_{a} = 0.1r_{vir} \simeq 7r_{d}$).  
Consequently the effect of radial orbit is not dramatic in Figure \ref{fig:Orbit}.
However it is clear that the radial orbit reduces the dark matter halo response.

Since it is difficult to simulate pure radial orbit halos in three-dimensional 
N-body simulation, we use implement one-dimensional simulation with a pure 
radial orbit halo.  The orbits are purely 
radial and the density profile of the simulated halo is proportional to $r^{-2}$ 
in central region. Figure \ref{fig:radial} shows the rotation curve of 
the simulated halo compared with the prediction of adiabatic contraction formula. 
The circular-orbit prediction overestimates the contraction by factor of two.

\section{Adiabatic contraction for typical CDM halos}
\label{sec:typical}

We consider a range of halo parameters and disk masses to explore the
general applicability of the circular-orbit adiabatic approximation.
Recent cosmological simulations show that velocities in the inner
region of dark matter halos is isotropic \citep{ENF98, CKK00, FM01,
DMS04}. Therefore we explore isotropic halos. 
Although an NFW halo model is currently accepted in CDM cosmology, 
some recent theoretical models \citep{MM02, MM04,
OB03, WK02} and observations \citep{Blok01} suggest that dark halos
may have cores.  Simulations for the core halo are also carried for
the three cases, with $c\equiv r_{vir}/r_{core}=15$, 12, and 7.5,
respectively.  In addition to the fiducial disk, $m_{d}=0.04M_{vir}$ and
$r_{d}=0.014 R_{vir}$, we consider a low-mass ($m_{d}=0.02M_{vir}$)
and a high-mass ($m_{d}=0.1M_{vir}$) disk.

\begin{figure}
\centerline{
\epsfxsize=1.0\columnwidth
\epsfbox{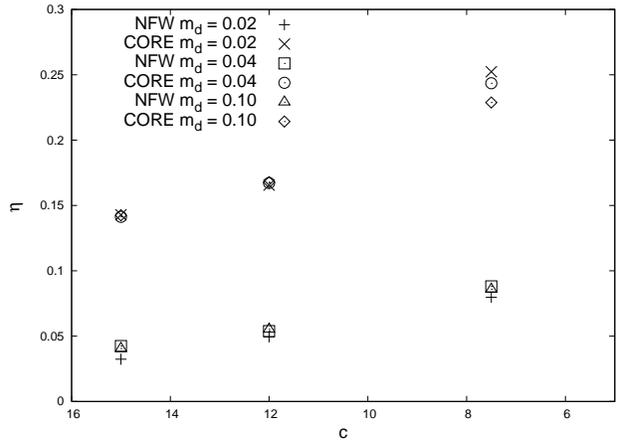}
}
\caption {
Relative difference ($\eta = \frac{|V_{adia}-V_{sim}|}{V_{sim}}$) between the rotation speed from the simulation
and from the adiabatic contract approximation at 2.2$R_{d}$ for c=15, 12, 
and 7.5 NFW and core halos. Three disks, $m_{d}=$0.1, 0.04,and 0.01 $M_{vir}$, 
are considered for each halo. 
The relative difference depends strongly
on the halo structure and the disk mass dependence is negligible.  }
\label{fig:ERROR}
\end{figure}

Figure \ref{fig:Vhalo} compares the post-formation rotation curves for
NFW halos and core halos with $c=15$ and $c=7.5$ with the
adiabatic contraction predictions.  The formula
overestimates the rotation velocity for these astronomically motivated
halo models although the discrepancy is modest.  Figure
\ref{fig:ERROR} quantifies the relative differences, $\eta\equiv\vert
V_{ad}-V_{sim}\vert/V_{sim}$, as a function of halo concentration $c$,
where $V_{sim}$ and $V_{ad}$ are the circular velocities at $r=2.2
R_d$ obtained from simulation and from the adiabatic contraction (eq.
\ref{eq_ad}), respectively.  The discrepancy is 4\% for NFW model with
$c=15$ and increases to 8\% for $c=7.5$.  The discrepancy
increases with disk mass but the dependence is weak.  The $c=7.5$ case
is a low value for galaxy halos in the current CDM model, therefore,
these results show that the adiabatic contraction approximation
remains good for isotropic NFW halos.  For core halos, the discrepancy
is 14\% for $c=15$ and increases to 23\% for $c=7.5$ with weak
dependence on $m_d$.

\citet{Gnedin04} improved the agreement between the adiabatic 
contraction approximation and the simulation result using a 
modified version of the original adiabatic contraction 
approximation including the effect of non-circular orbits.
In this model, the adiabatic invariant $M(r)r$ is replaced by 
$M(\overline{r})r$, where $\overline{r} = Ar^{w}$. The authors 
suggested  $A \approx 0.85 \pm 0.05$ and $w \approx 0.8 \pm 0.02$ 
based on cosmological dark matter halo simulations. 
As comparison, we show in Figure \ref{fig:Vhalo} 
the rotation curves obtained from this modified adiabatic 
contraction approximation, together with those obtained from the 
simulation and from the original adiabatic contraction approximation. 
As one can see, the modified adiabatic contraction approximation 
agrees with the simulation results for NFW halos. 
For halos with a constant density core, there is clear discrepancy 
between the model and simulation, although the modified model 
is better than the original adiabatic contraction model. 
Although the \citet{Gnedin04} model takes the effect of non-circular 
orbits into account, their fitting formula, $\overline{r} = Ar^{w}$, 
is not based on the orbital structure. It simply reduces 
the halo contraction using fitting formula with empirically suggested 
fitting parameters. Therefore the \citet{Gnedin04} model estimation for 
core halo cases with suggested fitting parameters does not work as well 
as one for NFW halo cases. For core halo, different fitting parameters 
and further modification in fitting formula are required.

\begin{figure}
\centerline{
\epsfxsize=1.0\columnwidth
\epsfbox{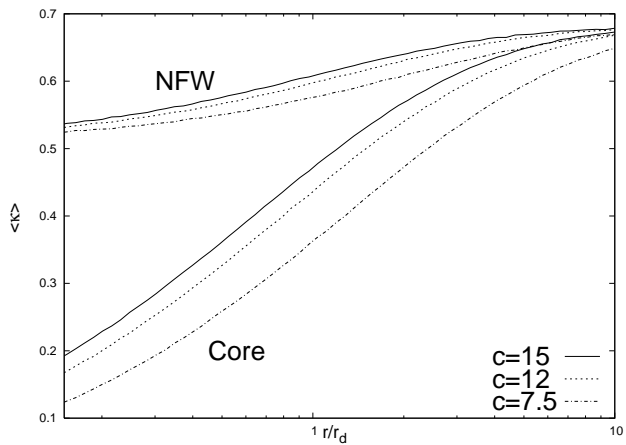}
}
\caption { The mean relative angular momentum per orbit, $\kappa =
  J/J_{max}(E)$, is averaged in radial bins for both the NFW and core
  halo models and three concentrations. The top three lines represent
  NFW halo models and bottom three lines represent core halo models.  
  The solid lines show c=15 NFW and core halo models, the dotted lines 
  shows c=12 halo models, and dot-dashed lines shows c=7.5 halo models.
  The fraction of eccentric orbits increases dramatically in core halos. 
}
\label{fig:kappa}
\end{figure}

Figure \ref{fig:ERROR} shows that the adiabatic contraction approximation 
is better for high-concentration halos.
This trend is explained by the distribution of orbits.
We describe the shape of an orbit by the ratio of the angular momentum 
to the angular momentum of a circular orbit at a fixed energy 
$\kappa = J/J_{max}(E)$.  The orbit with $\kappa=0$
($\kappa=1$) is radial (circular).  Figure \ref{fig:kappa} describes
the ensemble average of this ratio for isotropic halo in radial bins
$\langle\kappa\rangle(r)$.  At a fixed energy, the mean value of
$\kappa$ is 2/3 for comparison.  

Figure \ref{fig:kappa} shows that an isotropic NFW halo has more 
low-eccentricity orbits at fixed radius than an isotropic core halo. 
This can be understood as follows. The density at a given radius in a 
halo is contributed by particles on different orbits. In the inner region 
of a halo, orbits with lower energies are more circular, while those 
with higher energies are more radial. Assuming isotropic velocity 
dispersion, one can show that the energy distribution is flatter in a core 
halo than in a cuspy halo.
Consequently, for isotropic velocity 
distribution a core halo requires more high-eccentricity orbits
than an NFW halo.

\section{Summary}
\label{sec:summary}

We study the accuracy of the circular-orbit adiabatic approximation
\citep{blumenthal86} in predicting halo contraction due to disk
formation and provide a physical explanation for the deserved trends.  
We consider: (1) variation in the accretion time scale; (2) variation in
the accreted disk mass; (3) variation in the central concentration of
both cuspy and core halo models; and (4) variation in the velocity
isotropy.  The circular orbit adiabatic contraction approximation is
acceptable over a wide range of astronomically interesting
parameters. The relative change in the rotation curve value between
the simulation and circular orbit approximation at 2.2 disk scale
lengths, $\eta$, is less than 23\% for our entire range of realistic
parameters.  We find that the disk growth is still slower than dark
matter halo dynamical time in the vicinity of the disk and therefore
the adiabatic approximation is maintained.  The value of $\eta$
depends only weakly on the fraction of accreted baryon mass, and
therefore, the dependence on halo concentration cannot be explained by
disk dominance in less concentrated halos.  However, $\eta$ is
strongly correlated with the fraction of eccentric orbits in the
distribution. The steeper the cusp, the larger fraction of more
circular orbits are required at {\em fixed} radius, and this supports
the circular orbit approximation.  Although the adiabatic
concentration approximation overestimates the response of dark matter
halos, as long as the dark matter halo has central cusp and isotropic
velocity distribution the overestimation is negligible.

  Our results have important implications for the formation of 
disk galaxies in the CDM scenario. In comparing theory with the 
observed Tully-Fisher relation, one usually uses the peak rotation 
velocity of galaxy disks to represent the observed rotation 
velocities \citep{MMW98}. However, as shown in \citet{MM00}, 
if dark matter halos respond to the disk growth according to the  
adiabatic contraction model, current CDM model predicts a 
Tully-Fisher relation that has a much too low zero-point (i.e. 
galaxies are too faint for a given peak rotation velocity). 
In order to match the observed Tully-Fisher zero-point, one has to 
assume that disk growth does not cause any contraction  in dark 
matter halos at all \citep[e.g.]{croton06}. This assumption is not supported 
by our results, which show that adiabatic contraction approximation
works reasonably well for CDM halos over a wide range of situations. 
However, it should be pointed out that there are other astrophysical 
processes, such as mergers \citep{Dekel03,BM04}, dynamical heating by 
substructures \citep{EZ04}, halo pre-processing \citep{MM04}, and 
resonance dynamics \citep{WK02}, that may modify halo structures
but are not included in the models considered here. Unfortunately, 
the importance of these processes are not well understood at 
the present.

\section*{Acknowledgements}
We would like to thank our referee, Joel Primack, for many comments 
and suggestions that improved this paper.
We would like to thank Neal Katz for useful discussions.
JHC and YL thank Dusan Kere\v{s} and Yicheng Guo for reading our manuscript.
MDW acknowledges the support of NASA ATP NAG5-12038.


\appendix
\section{One-dimensional hydrodynamic simulations}

We use the Lagrangian based one-dimensional hydrodynamic code described in
\citet{lu06} to simulate formation of a halo from an initial perturbation 
at a high redshift $z_i$. We assumes the halo mass accretion histories
proposed by \citet{We02} 
\begin{equation}
M(a)=M_0 \exp\left[-2 a_c \left({a_0\over a}-1\right)\right],
\end{equation}
where $a$ is the expansion scale factor, $a_c$ is the scale factor
corresponding to the formation time of the halo, and $M_0$ is mass of the
halo at the observation time $a_0$.
In this function, $a_c$ is the only free
parameter to characterize the shape of a mass accretion history.
Reader may refer to \citet{lu06} for detailed description on making the
initial conditions given the mass accretion history. We choose $z_i=200$, 
and the initial temperature of gas shells is set to be the CMB temperature 
at this epoch.

The simulation has both dark matter shells and gas shells.
The gas initially follows the distribution of the dark matter but evolves
differently from the dark matter due to hydrodynamics.
We use the Lagrangian finite-difference scheme
to follow the evolution of the shells. The numerical treatment is same as
what is described in \citet{tw95}. To avoid numerical instability due to
dark matter shell-crossing, the mass of each dark matter shell is
chosen to be much smaller than that of a gas shell.
The baryon fraction is fixed at $f_b=0.17$.  
The chemical abundance is assumed to be primordial.  The
radiative cooling function proposed by \citet{katz96} is implemented
in the simulations.

When a gas shell cools to a temperature, below $10^4$K,
the gas in the shell is considered to be cold. Since we do not include
any cooling processes below this temperature, the cold gas is assumed to
retain a temperature of $10^4$K until it flows into the center of
the halo. At this point, the gas joins a central exponential disk
with a scalelength $r_d={0.05\over \sqrt2} r_{\rm vir}$, where
$r_{\rm vir}$ is the virial radius of the halo at $z=0$. The cold gas
disk is assumed to be a rigid object, and its gravity is included
in the subsequent evolution of other mass shells. At any given time, the
gravitational acceleration of a shell at radius $r_i$ is given by
\begin{equation}
g_i=H_0^2\Omega_{\Lambda}r_i - {GM(<r_i)r_i \over (r_i^2+r_\alpha^2)^{3/2}},
\end{equation}
where $H_0$ is the Hubble's constant at the present time, $\Omega_{\Lambda}$ is
the density parameter of the cosmological constant, $M(<r_i)$ is the total
mass (dark matter, gas and exponential disk) enclosed by $r_i$, and $\alpha$ is a softening
length taken to be 0.0005 of the virial radius of the halo at the present time.
This scale is much smaller than any scale of interest. In the
simulations, the density parameter of the non-relativistic
matter and of the cosmological constant are $\Omega_M=0.3$ and
$\Omega_{\Lambda}=0.7$. and the Hubble's constant is
$H_0=100{\rm km\,s^{-1}\,Mpc^{-1}}$.

\end{document}